# Development of Disciplined Interpretation Using Computational Modeling in the Elementary Science Classroom

Amy Voss Farris & Amanda Catherine Dickes, Vanderbilt University, USA
Pratim Sengupta, University of Calgary, Canada
Email: amy.farris@vanderibilt.edu, amanda.c.dickes@vanderbilt.edu, pratim.sengupta@ucalgary.ca

**Abstract:** Studies of scientists building models show that the development of scientific models involves a great deal of subjectivity. However, science as experienced in school settings typically emphasizes an overly objective and rationalistic view. In this paper, we argue for focusing on the development of *disciplined interpretation* as an epistemic and representational practice that progressively deepens students' computational modeling in science by valuing, rather than deemphasizing, the subjective nature of the *experience* of modeling. We report results from a study in which fourth grade children engaged in computational modeling throughout the academic year. We present three salient themes that characterize the development of students' disciplined interpretations in terms of their development of computational modeling as a way of seeing and doing science.

## Introduction

The development of a modeling-based epistemology in science is central to the development of scientific literacy (Nersessian, 2008; Lehrer, 2009). Models are fundamentally analogical forms (Giere, 1988), and studies of scientists building models show that the development of scientific models involves a great deal of subjectivity, and in many cases, is deeply intertwined with personal experiences and disciplined sensibilities about interpreting evidence (Keller, 1984; Ochs, Gonzales & Jacoby, 1996). However, science as experienced in school settings typically emphasizes an overly objective and rationalistic view (Lemke, 2001). This is particularly relevant for studies that involve learning science as modeling, because modeling is an act of design (Lehrer, 2009), and therefore, deeply interweaves "knowing" and "action" (Schön, 1995). This interweaving, as Schön (1995) argued, is deeply tied to subjectivities such as learning to see things from the perspectives of others, and engaging in reflective conversations with the situation. This is a far cry from common uses of computational modeling in the science classroom, which has traditionally followed the grossly "linear approximation" of teaching correct concepts through guided algorithmic refinement (e.g., White & Frederiksen, 1998; Booler & van Jooligen, 2013). Such images, grounded in technical rationality (Schön, 1995), leave out the development of necessary subjectivities, and this is the issue we address in this paper.

In this paper, building on Daston & Galison's (2007) notion of "trained judgment", we argue for focusing on the development of *disciplined interpretation* as an epistemic and representational practice that progressively deepens students' computational modeling expertise by valuing, rather than deemphasizing the subjective nature of the experience of modeling. We report results from a study in which fourth grade children engaged in computational modeling by iteratively creating, presenting and evaluating their mathematical measures and computational models of motion and ecology throughout the academic year. In this paper, we will focus on their models of motion and investigate how they develop progressively more mathematically and computationally refined representations of motion as a process of continuous change. We present three salient themes that characterize the development of students' disciplined interpretations in terms of their development of computational modeling as a way of seeing and doing science: (1) the intertwined nature of designing models as communicative forms and developing deeper interpretations of numerical data; (2) progressive development of criteria for what counts as a "good" measure, by extending the tools of modeling beyond computational media; and (3) the shift from more normative and canonical forms of mathematization to invented forms of simulations as equally "accurate" and communicative models.

## Theoretical Background

Science studies scholars have argued that the development of disciplined interpretation is central to the production of scientific knowledge. There is always a gap between scientific representations and reality; scientific models are by their very nature non-veridical designed artifacts that are deeply influenced by the technological infrastructure used for inquiry and representation, as well as the purpose of representation (Galison, 1996; Daston and Galison, 2007). These authors have argued that the epistemic stance of scientific work shifted from a falsely "objectivist" stance to "trained judgment". This was evident in their comparison between the 19th century introduction of photographic technology where the machinic nature of photography created an impression that scientists could "get out of the way," and let the photograph produce what became perceived as bare, un-interpreted, objective "facts." In contrast, beginning in the early to mid-twentieth century, with the advent of the printing press that in turn widened the audience for scientific works such as atlases, the production of scientific images became necessarily more interpretive on the part of the scientist, with a clear goal of *enhancing the communicativity* of the images (Daston & Galison, 2007). This represents another moment in the history of objectivity: *trained judgment.* We further posit that the advent of computing as a key mode and medium of scientific inquiry further amplifies this epistemic stance. For example, a recent ethnographic study of scientific work in a biomedical engineering lab illustrates how the inherent interdisciplinarity of the practice of computational modeling results in new conceptual innovations, often by bridging the gap between disciplinary perspectives, as well as between theorization, dynamic visualization and experimental work (Chandrasekharan & Nersessian, 2014).

How can disciplined sensibilities about modeling develop, especially in the context of computational modeling, in an elementary science classroom? This is the central concern of this paper. We posit that answering this question involves two key issues: The first issue involves finding a suitable paradigm of computing that is intuitive and generative for young children. Scholars have argued that a particular form of computation—agent-based computation—can serve as an effective pedagogical approach that can help children bootstrap their own pre-instructional ideas and representational competencies in order to develop scientific expertise through modeling (Papert, 1980; Sherin, diSessa & Hammer, 1993; Sengupta, Kinnebrew, Basu, Biswas, and Clark, 2013). In agent-based computation, users construct programs by providing simple rules to a computational object or agent, such as a LOGO turtle, which then enacts the rules through movement in computational space. Programming the agent involves thinking like it, which enables the learner to engage in embodied and intuitive reasoning (Papert, 1980; Danish, 2014). In agent-based models (ABMs), simple, agent-level actions are repeated over time (in the case of generating continuous movement from discrete actions) and/or across multiple agents (e.g., in ecological phenomena). There is ample evidence in the literature that ABMs can support the development of representational competence in children (Sherin, diSessa & Hammer, 1993; Sengupta et al., 2015), and that pedagogies that emphasize the creation of representational conventions that can be understood by others is a kind of "selective pressure" that brings forth iterative representational innovations by the students (Enyedy, 2005). We therefore adopted agent-based modeling and programming as the *medium* of computation that children engage with in our study.

Given the rather short duration of most education research studies and the limited involvement of the teacher in the design and implementation of what happens during class, little is known about what happens when computational modeling practices develop as a long-term practice in the science classroom over an entire academic year. Herein lies the second issue: *we posit that disciplined interpretation is one such aspect of development that fundamentally involves long-term engagement of learners with the practice of modeling.* While some scholars have shown that children can indeed engage deeply with agent-based modeling to learn science over shorter durations ranging from a few days to a couple of weeks (e.g., Danish, 2014), we posit that a long-term focus is essential because scientific modeling is itself a long-term and complex endeavor in professional practice. It comprises complex interactions among theories, material means, the phenomenon of interest, the representational infrastructure, as well as the social contexts that shape these interactions, usually over a long period of time—often spanning several years (Pickering, 1995). A longer-term focus in the science classroom can provide insights into how students develop fairly stable and sophisticated *disciplinary dispositions* that involve an interplay between children's intuitions, interpretations and actions on the world in order to progressively symbolize and refine the representations for scientific modeling (Lehrer, 2009). What does the development of children's disciplinary dispositions look like when computation becomes a medium for scientific modeling throughout the academic year? In this paper, our goal is to provide illustrative cases, selected carefully from a year-long study that will illuminate key characteristics, and different forms of manifestations of a particular form of disciplinary dispositions that is often neglected in the science education literature—disciplined interpretation.

## Methods

### The ViMAP Modeling Environment
The modeling platform we used is ViMAP (Sengupta et al., 2015). ViMAP is an agent-based visual programming language that uses NetLogo (Wilensky, 1999) as its simulation engine. In ViMAP, users construct programs using a drag-and-drop interface to control the behaviors of one or more computational agents. ViMAP programming primitives include domain-specific and domain-general commands as well as a "grapher" with multiple graphing windows, which allows users to design mathematical measures and compare across measures of different agent- and class-level variables. Figure 1 shows the programming interface and the graphing interface.

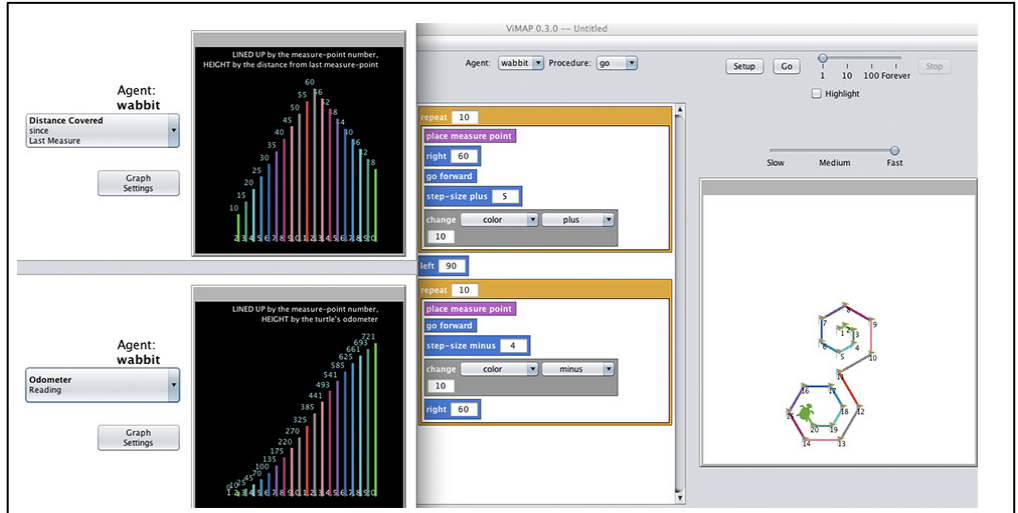

Figure 1. Screenshot of the ViMAP Modeling Environment (www.vimapk12.net).

## Participants, Setting & Data Collection

The data that we have collected is in the context of a public school fourth grade classroom (ages 9 and 10) in an urban southeastern city. The study is a design study (Cobb et al., 2003) in which we worked in partnership with the classroom teacher to integrate agent-based programming and modeling within the existing math and science curriculum. Students carried out investigations of natural phenomena in kinematics and in ecology in modeling cycles that include modeling in ViMAP. Twenty-one students and their teacher participated in the classroom work, which was embedded in the regular curriculum. The teacher and the research team co-planned the activities based on the students' progress and the teacher's plans across the curriculum. During class time, the teacher played the primary teaching role, often adjusting the plans to meet the emerging instructional opportunities as plans were enacted. Two graduate-student members of the research team collected data. Ninety-five percent of students who attend the school are eligible for free and reduced-price lunch. The ethnoracial and gender composition of the student population in this classroom was as follows: African-American: 19, Latino: 1; Somali: 1; Male: 11, Female: 10. A sequence of the learning activities is shown in Table 1. In this paper, we only report the analysis of all modeling activities from October 14 until February 3. The research approach was both microgenetic and sociogenetic, because our goal is to understand changes in student thinking and how these ideas are shared and taken up in the larger class community. The forms of data collection include video records of all classes, detailed field notes for each day, collection of all of students' models and non-computational artifacts, as well as interviews with student groups and individual students.

Table 1: Sequence of activities (Analysis reported in this paper through Feb 3).

| | | |
|---|---|---|
| Observations, pre-assessment, and interviews | Aug. 11 – Sept. 8 | Researchers conduct observations, preliminary interviews with all students in the class |
| Survival Kits Geometry Unit | Sept. 9 – Oct. 2 | Intro to ViMAP programming and modeling; Turtle geometry, centered around learning goals in perimeter, area, and angles of polygons; model sharing and revision |
| "Constant Speed" Robots | Oct. 14 – Nov. 20 | Students develop understanding of speed as a rate of the distance traveled in a unit of time, including cycles of model sharing and revision; students used both ViMAP and physical modeling |
| Constant Acceleration and Gravity | Nov. 25 – Feb. 3 | Students find ways to measure and model continuous changes in speed, using acceleration down a ramp and free fall as contexts; students used ViMAP, video analysis and physical modeling |
| Friction | Feb. 5 – Mar. 31 | Students model processes of "slowing down" for Matchbox cars on different surfaces; students used both ViMAP and physical |

|  |  | modeling |
|---|---|---|
| Interviews | Apr. 7 – Apr. 28 | Mid-year interviews with all students |
| Modeling Ant Colonies | May 6 – May 13 | Students model ant foraging, reproduction and predation in ant colonies in an embodied modeling activity, followed by programming in ViMAP-Ants. Students share and refine their models with 8th grade mentors. |
| Post-Assessment | May 14 - May 19 | End-of-year assessment and focus group |

## The Role of the Teacher & Researcher-Teacher Partnership

Teaching played a significant role in this study (Sengupta et al, 2015). Although a detailed analysis of teaching is outside the scope of this paper, it is important to note that the teacher reframed programming as a medium and activities for designing mathematical measures (i.e., units of measurement and graphs) of motion. The teacher co-designed (along with the researchers) and implemented learning activities that supported the interpretation and construction of mathematical measures using ViMAP as a way to explain a real-life phenomenon involving motion (e.g., walking and running). In these activities, she maintained an emphasis on connecting modeling in ViMAP to relevant out-of-computer modeling experiences, such as embodied and physical modeling activities. Furthermore, she created a culture for sharing and critiquing peer models, that is, their ViMAP programs, simulations and graphs. In this process, students began developing criteria for what features of their models would be worth sharing: the emphasis on *communicativity* acted as a selective pressure for model improvement; and, the class as a whole, normatively, developed criteria for what would count as a "good" computational model. These criteria originated in teacher-led class discussions as socially defined (voted by popular choice), but over time, became progressively more grounded in students' mathematical explanations of relevant aspects of their ViMAP simulations. This led students to use progressively more sophisticated computational abstractions, such as loops and conditionals, in order to make their models predictive (Sengupta et al., 2015).

## Analytic Approach

We conducted a thematic analysis (Miles & Huberman, 2004) in order to identify key forms of disciplined interpretations that learners developed during the phase of modeling motion. A theme captures something important about the data in relation to the research question, and represents some level of patterned response or meaning within the data set. In our study, at the highest level, each theme represents an interpretive judgment. Each theme, in turn, consists of sub-themes, which are sets of relevant *representational moves,* that is, actions undertaken by the learners that involve the creation, and/or editing of computational programs and other related representations, and *epistemic moves* (e.g., arguments about the validity or significance of certain representations). Over time, these representational and epistemic moves constitute, or lead to the development of an *interpretive judgment* (e.g., what counts as a "typical" measurement; what counts as a "good video"). These interpretive judgments developed through progressive refinement of models and moving back and forth across tangible, diagrammatic, and computational models of motion. Therefore, besides the learners' subjectivities, the judgments themselves are inextricably tied to the media involved in modeling, in addition to mathematical and physical ideas, and computational abstractions.

# Findings

## Interpreting numerical data & the emergence of ideas about error

In the first modeling cycle, the teacher wanted to design a context for students to define constant speed motion in terms of distances traveled per unit of time. The researchers initially wanted to use verbal descriptions of motion, as the best approximations of "constant speed" we would require motors. The teacher, however, insisted on introducing physical objects, both computational and non-computational, as part of the modeling activity. Her goal was to "make things concrete" i.e., to transform the modeling activity from a virtual and conceptual one into a lived-in experience for her students. The students were therefore provided with Lego Mindstorms NXT robots programmed to move at a constant speed. We also provided stopwatches, adhesive Post-it flags, and seamstress-style measuring tapes, and asked students to measure the distances traveled in regular intervals of time. Students coordinated the placement of position-marking flags with a stopwatch in order to come up with mathematical measures and explanations of robots moving at constant speed. In Figure 1, a student is measuring the distance between flags to find the distance traveled in each three-second interval.

Students used these measurements to create computational models of the motion in ViMAP. In their initial models, each group's measurements for the distances traveled in each three-second interval were non-uniform due to challenges inherent in the act of measurement. However, none of the students problematized their data by considering the limitations of the devices for measurement or the imperfect coordination of the placement of flags with the stopwatch. Students thought of themselves as workers following an unquestionable procedure. Some of them argued some about issues of fidelity *during* the measurement activity. However, once data collection was complete, no one thought critically about the potential problems in her or his data even though for most students, the data showed a wide variation in the measurements. Upon noticing this, the instructors designed an activity in which students watched a video of one group carrying out their measurement and data collection, and as a class, critiqued their work as shown in the video. The class replayed the video several times in order to notice and reflect upon successes and breakdowns in measurement. The ensuing discussion led to the first student talk about error: "It's not that the robot was moving differently, it's that *we* were making mistakes!" Making sense of the data with a focus on analyzing the *lived experience* of designing the measure helped students interpret data as *designed measures*, not independent of the challenges of measurement. This was particularly evident in their noticings of the various sources of error, as they replayed and re-analyzed the video. For example, some groups noticed that longer measurements of displacement were often coupled with shorter measurements, indicating that the timing of placement of the measure-flag shared by those measurements was likely off. Some groups also discovered errors due to misreading the measuring tape, and in some cases, due to the sticky flags being unintentionally moved by getting stuck to students' shoes..

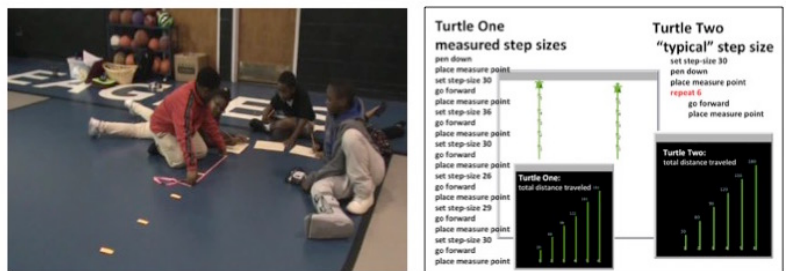

Figure 2. Measuring constant speed using adhesive paper flags as "measure flags" (left) and one student's model of measured and "typical" step-sizes (right).

We then asked students to review their measurements in order to determine what they believed was a "typical" distance measurement for their robot to travel in three seconds. The teacher welcomed this as an opportunity to connect the modeling activity with learning about measures of central tendency in their math curriculum. The students iterated upon their existing computational models using a second computational agent in the same simulation to represent the motion according to their "typical" values (mean, median, or mode) for speed. Figure 2 shows one student's model: her "measure-points" in the first-iteration were the following distances apart, measured in inches: 30, 36, 30, 26, 39, 30. Her second-iteration data shows six uniform measurements of 30 inches each. For each turtle, students programmed ViMAP to generate two graphs: one showing the value of each measurement (not shown), and another producing the total distance traveled by the robot. In the example shown in Figure 2, the graphs of the total distance traveled by the agent show a total distance of 191 inches for the measured data, and 180 inches for the adjusted, or "typical" data. Students also recognized that computationally, the typical model could be expressed as a loop, which the students and teacher appreciated as a more succinct program. A second affordance of the program for Iteration 2 is that the number of repeats can be changed to simulate the robot traveling for a longer period of time at the same speed.

In sum, making mathematical meaning of motion as processes of time-based change required the generation of and coordination among different *representational moves*, involving multiple forms of digital and paper-based, discrete-mathematical representations of the phenomena under study. Annotating video and photographic images in order to communicate and argue for the number of loops needed in their programs became a viable but emergent method for connecting among the representations, and can also be regarded as *epistemic moves* that grounded these representations within the disciplinary concepts. Students' agency and involvement in creating connections across representations for the purpose of making meaning represents a key practice in model-based reasoning. Epistemologically, the connections among representations were a shared unknown, and it was up to the members of the class to come up with and refine generative ways to see, quantify, and model salient aspects of motion.

### What counts as a "good" video for measuring acceleration?

Following students' refinement of their descriptions of constant speed and (average) speed on inclines of varying steepness, we began to work on developing descriptions of acceleration. We provided clear acrylic tracks, marbles, and Lego bricks to build supports for the top and bottom of the tracks, as well as stopwatches and adhesive paper flags, in order to begin to measure acceleration. Students' initial descriptions neglected processes of continuous change: the marble was "slow" at the top of the ramp, and "fast" at the bottom. When asked to measure how speed was changing, students tried to reapply their method with the robots: they attempted to place flags at equal intervals of time, but they soon decided this was too difficult: the motion was *too fast* for the method used when measuring the speed of the (slower) robots. As a potential solution to this problem, the instructors introduced digital video as a new method for collecting and analyzing motion data. We made this design decision because authoring videos leverages out-of-school literacies for making and working with digital video, and, high frame-rate videos afford the possibility of slowing down recordings of motion that are otherwise too fast to measure.

Children's ideas of what counted as a "good" video for measuring acceleration changed dramatically between the first and second iterations of their video recording and subsequent analysis. An example of student work is shown in Figure 3. Initially, their videos followed the marble in an action perspective, but this made measurement impossible because there was no frame of reference from which to measure the distances traveled. After attempting to measure the acceleration of the marble using the first round of videos, and after classroom-wide discussions, the class developed a norm for what counts as a "good" video for measuring motion in a frame-by-frame analysis: the camera has to stay still and the field of view must show the whole motion. Central to this was the realization that the viewer should be able to see the marble and identify the exact frame at which it was released. Students iteratively developed, shared and critiqued several videos, and progressively refined their measures, and over time, developed measures using physical and material means that used discrete mathematical representations similar to their ViMAP turtle's "step-size". One common form of measure involved placing a flag on their computer screens to mark the position of the marble at regular intervals of video frames. The pattern of the increasing distances between successive flags visually represented acceleration as a continuous process of change (Figure 3). The students' *epistemic moves*, evident in the form of their explanations and concerns for what makes a video of motion useful for measuring speed represents the students' understanding of important aspects of motion from a disciplinary perspective. These measurements, only possible through an innovative use of video, consisting of a set of eminently *representational moves*, yet deeply interwoven with the *epistemic moves*, were then used to make computational models that further communicate the mathematical pattern of change, as discussed in the following section.

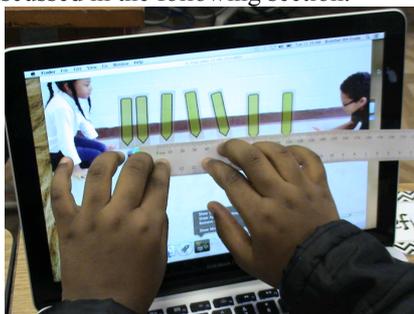

Figure 3. Adhesive flags mark the position of the marble at a regular interval of 10 frames.

### Expanding views of "accuracy" to create visually communicative models

In the first two modeling cycles (Oct. – Dec.), conceptual understanding of the unit of "speed" played an important role in their state-mandated science curriculum, and the teacher emphasized distance traveled by the ViMAP turtle in one "step" as the representation for speed. During this phase of the study, the *goodness* of ViMAP models as a representation of motion was normatively evaluated by the class based on the match between the speed vs. time graph ("distance covered since last measure"), and the speed data that the model was designed to represent. Given that the graphs made the pattern of change explicit in these models, the students came to see graphs as the primary communicative devices, and the turtle enactment (i.e., the geometric shape generated by the turtle commands) was seen as merely the means to generate the graph. This was evident in multiple episodes of students' sharing and presenting their models with the class. Over time, especially during Jan. – Feb., the instructors began to encourage students to further explore the ViMAP commands library, so that they would begin to deepen their use of the programming language. The goal here was to prompt students to re-envision and re-design their models using newer turtle variables, so that they could make their turtle graphics less literal and more visually and mathematically communicative. As a result of this instructional push, during the third modeling cycle, all students began to take a more design-oriented approach to producing models that communicate the most important ideas, and variation in student models emerged. In terms of *representational moves*, all the students in the class expanded their use of *variables* by using new commands to represent mathematically the gradual change in speed using one or more of the following variables: rotation, contrasts of color, pen-width, or relative size of agents. This was, in turn, motivated by and inextricably related to an *epistemic move*, that is—the students' goals of making relevant features of the phenomenon (motion) more salient to the class during presentations in their ViMAP models. We illustrate this change with the work of one student, Darien. His first model of constant acceleration is shown in Figure 4 (left). The figure shows the inscription made by the agent as it executes the associated commands. The model increases the distance traveled by the agent by two step-size units with each step. However, the enactment itself is limited in communicating the regularity of the increase—one would need to look at Darien's code or at the graphs to understand the regularity of the change.

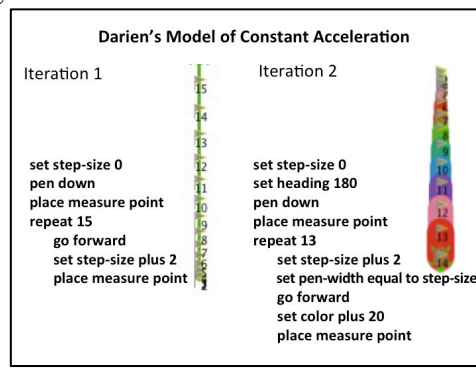

Figure 4. Darien's first (left) and final (right) models of acceleration.

In his final model (Figure 4, right), Darien added color changes to visually differentiate the individual steps of the agent, and co-varied the pen-width with the step size using the command *set <pen width> equal to <step-size>*. Darien presented his model to the class, describing that the increasing pen-width of the ViMAP turtle is intended to communicate that the ball is getting faster as it falls. When he was sharing his work with the class, students asked questions about the representational significance of different aspects of his model, such as "Why does it look like a baseball bat?", "Can we see the data", and pointing out redundancies in his code (an initial *set <pen-width>* command, that was being overridden by the co-variation command). An excerpt of talk from Darien's presentation is shown below:

Akia: Why do you have *set pen width* equal to *step size*?
Darien: To actually help the pen width [be] equal to the step size, so that way, the pen, the size will actually get bigger EV'RY STEP. [holds a hands slightly apart to show a space between, gesture beats and enlarges at syllables of ev-'ry step]
Teacher: So when it gets bigger, does that show that your speed is increasing or decreasing?
Darien: Increasing.

Darien wanted to show that pen-width was increasing with every step, and illustrates this with his hand movements as he speaks, beginning with his hands slightly apart, and enlarging the space on beats with the syllables "ev-'ry step". One could argue that Darien's initial model was more canonical, because it uses the commonly used representations of dot-traces and graphs. However, in his revised model, Darien's goal was to make the process of a steady increase in speed explicit *without* the use of graphs. This in turn led him to using an interpretive move that involved using a computationally more sophisticated data representation—co-variation—in order to link the visual appearance of his model (*pen-width*) to a variable that was significant in terms of representing the underlying physics (*step-size,* or speed).

## Conclusions and Implications

We have argued here that when children can engage in long-term, extended cycles of modeling using agent-based computational platforms (e.g., ViMAP) and complementary forms of physical modeling, they can begin to develop disciplinary dispositions and sensibilities pertaining to scientific modeling that parallel the more mature interpretive work of scientists (e.g., see Daston & Galison, 2007). The long-term nature of the study allowed students to connect representational experiences across modalities, including their computational representations with their lived experiences of designing measures in the real world with physical objects. Through such experiences, children came to view data as designed measures, and their views of what counts as a "good" model deepened significantly as they engaged in cycles of sharing and refining their models to be progressively more communicative. The emphasis on communicativity also led students to make deeper forays into programming and computational thinking.

Our work has implications for the praxis of computational modeling in the science classroom. There is now a growing body of literature that argues for the use of multiple and complementary forms of modeling in the classroom (e.g, Danish, 2014; Dickes et al., in press). In our study too, the students' representational and epistemic work were distributed across a range of computational and non-computational materials, using which they iteratively represented motion as a process continuous change. While modeling with ViMAP enabled the students to connect graphs of change over time to units of change (e.g., step-size), modeling with materials complemented this activity by enabling them to generate the phenomenon being modeled in the "real world", as well as to design the measure of change (e.g., step-size) using video analysis. It is also important to note that "making things concrete" using material forms was an instructional push initiated by the teacher, who co-designed these activities with the research team. We therefore believe that designing complementary forms of computational and non-computational modeling is critical for enabling teacher-adoption and appropriation of computational modeling in the K12 science curricula.

## Acknowledgement

This research was supported in part by an NSF Early CAREER Award (PI: Pratim Sengupta). All opinions are the authors' and not endorsed by supporting institutions.